\documentclass[reprint, amsmath,amssymb, aps, nofootinbib]{revtex4-2}
\usepackage{bm, physics,graphicx,float}

\begin{document}

\title{Observable and Unobservable in Quantum Mechanics}

\author{Marcello Poletti}
\email{epomops@gmail.com}
\affiliation{San Giovanni Bianco, Italy}
\date{\today}

\begin{abstract}

This work explores the connection between logical independence and the algebraic structure of quantum mechanics. Building on results by Brukner et al., it introduces the notion of \textit{onto-epistemic ignorance}: situations in which the truth of a proposition is not deducible due to an objective breakdown in the phenomenal chain that transmits information from a system A to a system B, rather than to any subjective lack of knowledge. It is shown that, under such conditions, the probabilities accessible to a real observer are necessarily conditioned by decidability and obey a non-commutative algebra, formally equivalent to the fundamental postulates of quantum mechanics.

\end{abstract}

\maketitle

\section{Introduction}
Classical Probability Theory (CPT) and Quantum Mechanics (QM) are two profoundly different algebraic frameworks that nevertheless yield similar outputs: real values in the interval \([0,1]\), representing the frequency or probability of a given physical phenomenon.

These two frameworks are incompatible, as definitively demonstrated by Bell \cite{Bell,Bell2,Bell3,Aspect, Cirillo, Bertlmann}. In particular, QM violates certain inequalities that are provable theorems within CPT. In 1981, Bell summarised his results in the celebrated article on Bertlmann’s socks\cite{Bell2}, presenting a particularly simple version of these inequalities, known as the Wigner–d’Espagnat inequality (WE):
\begin{equation}
p(A \wedge B) + p(\neg B \wedge C) \geq p(A \wedge C)
\end{equation}

The probability of both \(A\) and \(B\), plus the probability of both \(\neg B\) and \(C\), is greater than or equal to the probability of both \(A\) and \(C\). This inequality follows directly from a corresponding theorem in set theory, to which CPT is tightly connected via Kolmogorov’s formalisation\cite{Kolmogorov}:
\begin{equation}
\{A \cap B\} \cup \{B^c \cap C\} \supseteq \{A \cap C\}
\end{equation}

That is, the set of elements that are both \(A\) and \(B\), together with those that are both \(\neg B\) and \(C\), contains all elements that are both \(A\) and \(C\).

The proof is elementary: if \(x \in A \cap C\), then either \(x \in B\) (and thus \(x \in A \cap B\)), or \(x \in B^c\) (and thus \(x \in B^c \cap C\)).

On the other hand, in a Stern–Gerlach-type experiment, the quantum mechanical prediction is:
\begin{equation}\label{SG}
p(N_\alpha, S_\beta) = \frac{1}{2} \sin^2\left( \frac{\beta - \alpha}{2} \right)
\end{equation}

Here, \(p(N_\alpha, S_\beta)\) denotes the probability that an electron is detected in the “North” channel under a magnetic field oriented at angle \(\alpha\), and in the “South” channel under a field at angle \(\beta\).

The Wigner–d’Espagnat inequality would, in this case, imply:

\begin{equation}
p(N_0, S_{45}) + p(N_{45}, S_{90}) \geq p(N_0, S_{90})
\end{equation}

But evaluating these quantities using equation \ref{SG} yields:

\begin{equation}
\begin{cases}
p(N_0, S_{45}) + p(N_{45}, S_{90}) \approx 0.15 \\
p(N_0, S_{90}) \approx 0.25
\end{cases}
\end{equation}

QM violates CPT—and more than that, it contradicts predictions derived from a set-theoretic logic grounded in the law of the excluded middle. Does QM violate this principle?

The centuries-old debate surrounding the interpretation of the superposition principle—and more generally of QM—can largely be traced back to this point: within the set \(A \cap C\), there appears to exist an element that is neither an element of \(B\) nor an element of \(\neg B\). This tension emerges in popular expressions such as “The cat is neither alive nor dead,” or “The electron is neither here nor there.”

Such conceptual difficulties are also reflected in expressions like “quantum logic,” \cite{Birkhoff,VonNeumann} which seems to refer to an alternative logic that, in some sense, contradicts classical logic. Does QM violate logic? Is QM illogical?

The study of the relationship between logic and QM has been explored from many perspectives\cite{Benioff,Wheeler,Poletti,Poletti2,Poletti3,Poletti4,Szangolies, Landsman}; here, we resume this inquiry by drawing in particular on a series of works by Brukner et al.\cite{Brukner,Brukner2,Brukner3}, concerning the relation between logical indeterminacy and quantum randomness.

By “logical indeterminacy,” we refer to the impossibility of determining the truth or falsity of a proposition \(q\), given a set of premises \(p_1, p_2, \dots, p_n\). In elementary terms, from the premises “Socrates is a man” and “All men are mortal,” one can logically deduce “Socrates is mortal”; in contrast, the proposition “Socrates is Athenian” is independent of those premises and thus unprovable with respect to them.

Brukner’s work highlights an intriguing fact: the formalism of QM can be used to encode semantic relations among propositions. Specifically, given a set of premises, one obtains a state vector in an appropriate Hilbert space; a proposition whose truth value is not inferable from the premises corresponds to a quantum measurement with multiple and stochastic outcomes. The independence of a proposition with respect to a given set of axioms becomes experimentally testable by constructing appropriate quantum experiments.

These results suggest the possibility that quantum randomness may be a logical necessity rather than an ontological fact.

The aim of this work is to show that the relation between logical indeterminacy and quantum physics can be extended further. The presence of logical indeterminacy—in a specific form that will be defined as \textbf{onto-epistemic}—entails the necessity of a probabilistic framework that is formally equivalent to QM and thus, like QM, violates Bell-type inequalities. The problem of propositional independence requires that science adopt a probabilistic structure and, in certain contexts, to break with the theorems of classical probability theory.

\section{The Measurement Process}

In his celebrated \textit{Foundations}\cite{VonNeumann2008}, von Neumann devotes the final chapter to the \textit{Measuring Process}.\footnote{The italicised passages below are exact quotations from the 2008 edition.}
In this chapter, measurement is defined as a \textit{process of subjective perception}, irreducible to mechanics because it \textit{leads us into the intellectual inner life of the individual.}

Measurement is the process by which a physical state (for example, "the cat is on the table") is transformed into a mental state ("I know the cat is on the table) and finally into a proposition expressed in natural language ("The cat is on the table"). That is, measurement is the process that enables what may be called a Tarskian notion of truth, in which syntax and semantics converge: The statement "snow is white" is true if and only if snow is white.

The correspondence between reality and the assertion of states of reality is mediated by what von Neumann calls \textbf{principle of psycho-physical parallelism.} This principle ensures that \textit{it must be possible so to describe the extra-physical process of subjective perception as if it were in the reality of the physical world; i.e., to assign to its parts equivalent physical processes in the objective environment, in ordinary space}, thereby guaranteeing the very possibility of scientific knowledge—that is, again, of transforming reality into mental states, beliefs, theories, and so forth.
Von Neumann avoids turning the measurement problem into a theory of mind by noting that the psycho-physical parallelism can be cut at an arbitrary point. Consider the following scenario: in an experiment\footnote{Von Neumann describes a slightly different but analogous case.}, a screen lights up at a given spot—that is, a phosphor on the screen emits photons (or electromagnetic radiation, or something else); these photons reach the retina, which converts the incoming signals into electrochemical impulses that travel along the optic nerve and eventually reach deep regions of the brain, triggering a complex neurological reaction that, through largely unknown mechanisms, produces what we have called a mental state: "I know that spot on the screen is illuminated". Now, von Neumann observes that what is mechanically relevant is merely that the transmission occurs—that information propagates—even if we entirely ignore the workings of consciousness, even if we cut the phenomenal chain at the level of the retina, or earlier, thereby avoiding the need to incorporate into the theoretical system components so complex as to be formally unmanageable.

In the pages that follow, von Neumann proceeds to discuss in technical terms how the state vector propagates beyond the event we naïvely regard as the outcome of the experiment ("that phosphor lit up"), eventually encompassing the surrounding environment and transforming into what properly constitutes a measurement—that is, the \textit{subjective perception} corresponding to the establishment of the mental state "I know that phosphor lit up".

These profound observations by von Neumann suggest three considerations.
The first is that, although they were developed in the specific context of quantum mechanics, they in fact apply to any scientific theory. The scientific method, after all, rests on the comparison of two mental states: “I predicted that...” and “I measured...”. And again, even ignoring how such states are produced, it would still always be necessary to assess how the relevant information has propagated so as to reach the space-time neighbourhood of the observer.

Even in classical mechanics, a proposition such as "On that day, at that time, Jupiter will be in that position" is, strictly speaking, scientifically irrelevant, as it describes an ontological state—a Kantian noumenon\cite{Kant}—that cannot be measured unless accompanied by a psycho-physical transmission, a phenomenal chain, which allows a scientist on Earth to assert: "Yes, Jupiter is now in that position; the theory is confirmed."

The second consideration is that von Neumann’s cut allows the problem of psycho-physical parallelism to be reduced to a mechanical question, entirely disregarding the role of consciousness and, in a certain sense, even of measurement itself. That is, given any two physical systems A and B, the question arises of how one responds to the state of the other; that is, A can respond to some property of B only if there exists a phenomenal chain that conveys information about the state of B to A.

Finally, the third observation—which is the central point of this work—is the following: psycho-physical parallelism has historically been assumed as a given, and it ensures that the ontological states of classical mechanics are measurable. For this to hold, it is necessary that the phenomenal chain transmit information continuously and completely. But what would happen if this transmission were incomplete or discontinuous? What effects would this fact have on the theories and on the measurements performed by a sentient agent?

Epistemic ignorance is classically understood as a condition associated with a conscious agent, concerning the set of truths that the agent does or does not know, and it is remediable by acquiring additional information. However, there exists a second level of epistemic ignorance: we define \textbf{onto-epistemic ignorance} as the condition in which, in a system A, a proposition $p$ concerning a system B is unprovable not because of the agent’s epistemic limitations, but because of the absence of phenomenological chains capable of transferring information from B to A.

Consider an oscillating system $A$, for example, a body with a two-valued property—Up and Down—that changes over time.
Near this body, place a second body $B$, which responds to changes in $A$ by itself adopting the states Up and Down.

What is observed, from a Kantian perspective, is more properly that $A$ oscillates and emits phenomena associated with this oscillation (e.g., a magnetic field), and that $B$ does not react to changes in $A$ directly, but rather to the local variation in the phenomenological chain.

Now, suppose this phenomenological chain could be shielded—in other words, if no information about $A$’s state reaches B. In that case, $B$ would no longer react to
$A$. \textbf{Onto-epistemic} ignorance would then concern not merely the knowledge state of an agent, but the objective behaviour of physical systems.

A sentient agent—a scientist—enters the picture when one asks what kind of scientific theory, and in particular what kind of theory of measurement, such an agent
might construct when subjected to onto-epistemic ignorance.

As long as the scientist remains outside the shielding of the phenomena, they do not have— even in principle—any information that would allow them to determine the state of $A$. The proposition “$A$ is Down” is inaccessible. In this situation, the agent can only produce probabilistic predictions—even in the complete absence of deterministic chaos, even if the object under study were a simple, non-chaotic deterministic mechanism.

\footnote{The philosophy of quantum mechanics naturally also introduces the possibility that probability arises from an objective absence of the properties themselves, and that such properties become ontological features of the system only as a result of the measurement process (and of a non-local interaction). The aim of the present work, however, is precisely to show that this radical and problematic ontological position is not necessary. This framework is compatible with such a scenario; however, it does not require assuming that properties are not encoded in the object itself.}

Conversely, the scientist can “perform a measurement”—that is, enter the unshielded region, bring the body $B$, respond to the image formed on their retina by the position of $B$, which now in turn responds to the phenomena of $A$, and thereby establish the objective state of $A$. Finally, the scientist applies the scientific method by comparing their predictions to the results of the measurements.

Finally, it should be noted that the shield used in the previous thought experiments is merely a rhetorical device. Onto-epistemic ignorance arises, in general, whenever the phenomenal chain is incomplete or discontinuous, regardless of whether any physical shielding is present.

Before proceeding, a terminological clarification is in order. In particular, it is necessary to identify two terms that will designate, respectively, propositions whose truth value is determinable (because phenomenally accessible) and those that, by contrast, are inaccessible due to onto-epistemic limitations.

The terms \textit{decidable} and \textit{undecidable} would, from an intuitive standpoint, be a natural choice. However, their technical meaning in formal logic renders them unsuitable for operational use in the present context.\footnote{Several reviewers were very explicit on this point.} Their use would introduce significant conceptual ambiguities. Likewise, terms associated with Bell, such as \textit{speakable} and \textit{beable}, are now burdened with theoretical and historical connotations that would create analogous ambiguities.

Other authors have employed expressions such as \textit{manifestly true/false}, \textit{determinate/indeterminate}, or \textit{determinateness/indeterminateness}. Yet each of these options presents semantic difficulties: some imply a theory of mind, others suggest an ontological background that is not adopted here.

In what follows, the terms \textit{observable} and \textit{unobservable} will be used. Employed as adjectives, they should not unduly conflict with the noun \textit{observable} (indeed, they are intended precisely as its adjectival extension), and they appear to capture the definition of onto-epistemic ignorance with reasonable clarity and naturalness.

We shall therefore say that a proposition is \textit{observable} when its truth value is determinable by a local observer—that is, one not subject to onto-epistemic ignorance; conversely, a proposition will be said to be \textit{unobservable} when, although it concerns an objective system, it cannot be evaluated, within a given neighbourhood, due to a discontinuity in the phenomenal chain.

\section{Onto-Epistemic Probability}

In the previous section, two physically and conceptually distinct \textit{zones} emerged:  

\begin{itemize}
	\item Beyond the shielding, the proposition $p =$ \textit{``$A$ is Down''} is unobservable. We will denote this zone by $U$.
	\item Within the shielding, $p$ can be inferred; the phenomenal chain allows $B$ to respond to $A$, and the scientist to obtain the objective value of $p$. We will denote this zone by $O$.
\end{itemize}

The key point is that, in $U$, it is by construction impossible to measure $A$. The only way to measure $A$ is to shift to the $O$ perspective, where the proposition $\bar{p}$= \textit{``$p$ is observable''} holds true.  

In other words, there is no way—by construction—to empirically verify the proposition $p$ alone, but only the compound proposition \textit{``$p$ and $p$ is observable.''}  
This fact has an important consequence for probability theory, since the actual frequencies that can be measured in a laboratory are always—by construction—probabilities of \textit{$p$, given that $p$ is observable}.  

We will denote by $|p|$ the classical probability of $p$, and by $[p]$ the probability induced by onto-epistemic ignorance. We shall call this \textit{relative probability}, to emphasise that it depends not only on the truth value of $p$, but also on whether that truth value is observable or not within a given neighbourhood.  

Classical probability is defined in terms of set-theoretic measures that represent an absolute condition of truth or falsity. In contrast, for a real observer, the actual frequency values collected in the laboratory are necessarily subject to stronger constraints: the real measured frequencies are always conditional probabilities.  

Classical probability is governed by the joint probability theorem, from which, in particular, one obtains:
\begin{equation}\label{joint}
|p \land q| = |p||q|_p = |q||p|_q
\end{equation}
where $|q|_p$ and $|p|_q$ are the algebraic symbols associated with the \textit{probability of $q$, given $p$} and the \textit{probability of $p$, given $q$}, respectively.  

The probability induced by onto-epistemic ignorance can therefore be defined classically as:
\begin{equation}\label{relativeprobabilitypre}
[p] = \frac{|p \land \bar{p}|}{|\bar{p}|}
\end{equation}

From which, using equation \ref{joint}, we obtain:
\begin{equation}\label{relativeprobability}
[p] = |p|_{\bar{p}}
\end{equation}

The most evident consequence of \ref{relativeprobability} is that \ref{joint} becomes empirically inapplicable.  
However, even in a relative scenario, it is still possible to define, in a semi-classical manner, the concept of $[p]_q$, that is, the \textit{relative probability of $p$, given $q$}.  

Let $T$, $F$, and $U$ denote the conditions of "true and observable", "false and observable", and "unobservable", respectively. A pair of such symbols (e.g., $TT$, $TF$, ...) will represent the joint state of $p$ and $q$.  
The relative probability of $p$ can then be defined, in terms of these joint states, as:
\begin{equation*}
[p] = \frac{\|TT\| + \|TF\| + \|TU\|}{\|TT\| + \|TF\| + \|TU\| + \|FT\| + \|FF\| + \|FU\|}
\end{equation*}

And $[q]_p$ can be defined as the relative probability of $q$ in the states where $p$ is true and observable, that is:
\begin{equation}
[q]_p = \frac{\|TT\|}{\|TT\| + \|TF\|}
\end{equation}

From which it follows directly that:
\begin{equation}
[p][q]_p \neq [q][p]_q
\end{equation}

In other words, a scientist subject to onto-epistemic ignorance observes a form of non-commutativity that apparently violates equation \ref{joint}, and therefore a theorem of classical probability.

To make this deviation more tangible, consider the following example.  
Consider two entities, $X$ and $Y$ (balls, coins, etc.), each of which can exhibit two properties: colour (white W, black B) and orientation (up, down).  A shield filters out all information regarding objects in the "up" orientation. As a result, from the perspective of the observer, X and Y may appear as white or black, and either visible or invisible.

To represent visibility, we will use uppercase letters for visible objects ($W$, $B$) and lowercase letters for invisible ones ($w$, $b$).  
Each entity can therefore be, relative to an observer, in one of four states: $W$, $B$, $w$, $b$.  
The system $(X,Y)$ will thus have a total of $16$ possible combinations.  

For simplicity, we assume that any combination involving an invisible state has zero probability, except for the case in which $X$ is visible and $Y$ is invisible and white, i.e., $(X = W, Y = w)$.  
The system then reduces to five relevant configurations: $WW$, $Ww$, $WB$, $BW$, $BB$, with respective probabilities: $|WW|$, $|Ww|$, $|WB|$, $|BW|$, $|BB|$.  

Now consider the propositions $p=$ "$X$ is white" and $q= $ "$Y$ is white".  
In classical probability, we have:
\begin{equation}
\begin{array}{l}
	|p| = |WW| + |Ww| + |WB| \\
	|q| = |WW| + |Ww| + |BW| \\
\end{array}
\end{equation}

And the conditional probabilities are:
\begin{equation}
\begin{array}{l}
	|p|_q = \frac{|WW| + |Ww|}{|WW| + |Ww| + |BW|} \\
	|q|_p = \frac{|WW| + |Ww|}{|WW| + |Ww| + |WB|} \\
\end{array}
\end{equation}

From which it follows that:
\begin{equation}
|p||q|_p = |q||p|_q = |WW| + |Ww|
\end{equation}

However, in a phenomenal context, an observer can only measure visible states.  
The only invisible state allowed is $Ww$, in which $Y$ is invisible; thus, this configuration does not contribute to the empirically observable frequencies for $Y$.  

The empirical frequencies observed will therefore be:
\begin{equation}
\begin{cases}
	[p] = |WW| + |Ww| + |WB| \\
	[q] = \frac{|WW| + |BW|}{|WW| + |WB| + |BW| + |BB|} \\
\end{cases}
\end{equation}
And, for the corresponding relative conditional probabilities:
\begin{equation}
\begin{cases}
	[p]_q = \frac{|WW|}{|WW| + |BW|} \\
	[q]_p = \frac{|WW|}{|WW| + |WB|} \\
\end{cases}
\end{equation}
Letting $|Ww| = \epsilon$ and $|WW| = \Pi$, we obtain:
\begin{equation}
\begin{cases}
	|p||q|_p = |q||p|_q = \Pi + \epsilon \\
	[p][q]_p = \frac{|p|}{|p| - \epsilon} \cdot \Pi \\
	[q][p]_q = \frac{\Pi}{1 - \epsilon} \\
\end{cases}
\end{equation}

In general, therefore, $[p][q]_p \neq [q][p]_q$; and in the absence of invisibility ($\epsilon = 0$), i.e., in the absence of onto-epistemic ignorance, the three products coincide once again:
\begin{equation}
[p][q]_p = [q][p]_q = |p||q|_p = \Pi
\end{equation}

\section{Relative Probability and Quantum Mechanics}

In this section, we will show that the relative probability $[p]$ induces a complex, non-commutative algebra of projection operators—namely, the algebra defined by the fundamental postulates of quantum mechanics.  
By this we refer to the structural postulates of QM that define the algebraic framework: Hilbert spaces, state vectors, Hermitian operators, and the Born rule—excluding the more properly dynamical postulate, the Schrödinger equation, which is understood here as the quantum formulation, within the algebra defined by the preceding postulates, of the Hamiltonian formalism.  

For simplicity, the demonstration will concern only observables with two possible values ($0$–$1$, true–false, yes–no, etc.); that is, the subset of QM known as \textit{quantum logic}.  

The demonstration will proceed in three steps:

\begin{enumerate}
	\item First, CTP and QM will be brought closer by reformulating CTP within a pseudo-quantum algebra, composed of vector spaces, state vectors, and linear operators.
	\item The resulting algebra will then be used as a stepping stone to provide an analogous operator-based representation of relative probability.
	\item Finally, the identity between this algebra and QM will be established.
\end{enumerate}

\subsection{Geometric Representation of Classical Probability Theory}

Equation \ref{joint}, by introducing a new algebraic symbol, highlights a typical semantic issue in classical probability theory. The probability of “\( p \wedge q \)” is not an algebraic function of the probabilities of \( p \) and \( q \), but depends, in general, on the semantic relationship between the propositions.  
Conversely, given the probabilities \( |p \wedge q| \), \( |p \wedge \neg q| \), \( |\neg p \wedge q| \), and \( |\neg p \wedge \neg q| \), it is possible to reconstruct the probabilities of \( p \) and \( q \). In particular:

\begin{equation}
|p| = |p \wedge q| + |p \wedge \neg q|
\end{equation}

In other words, the semantic relationships between \( p \) and \( q \) are captured by the quadruple:

\begin{equation}
\{ |p \wedge q|,\ |p \wedge \neg q|,\ |\neg p \wedge q|,\ |\neg p \wedge \neg q| \}
\end{equation}

A powerful geometric representation of these objects can be obtained by considering the vectors:

\begin{equation*}
|s\rangle = \left( \pm \sqrt{|p \wedge q|},\ \pm \sqrt{|p \wedge \neg q|},\ \pm \sqrt{|\neg p \wedge q|},\ \pm \sqrt{|\neg p \wedge \neg q|} \right)
\end{equation*}

These vectors are subject to a form of the law of the excluded middle\footnote{Since we are working over \( \mathbb{R} \), there are no complex conjugates, so \( \langle s| \) can simply be considered equivalent to \( |s\rangle \).}:

\begin{equation}
\langle s|s \rangle = 1
\end{equation}

To the propositions \( p \) and \( q \), one can now associate diagonal projectors that extract only the components of \( |s\rangle \) in which \( p \) (respectively \( q \)) is true:

\begin{equation*}
P = \begin{pmatrix}
	1 & 0 & 0 & 0 \\
	0 & 1 & 0 & 0 \\
	0 & 0 & 0 & 0 \\
	0 & 0 & 0 & 0 \\
\end{pmatrix}
\end{equation*}
\begin{equation*}
Q = \begin{pmatrix}
	1 & 0 & 0 & 0 \\
	0 & 0 & 0 & 0 \\
	0 & 0 & 1 & 0 \\
	0 & 0 & 0 & 0 \\
\end{pmatrix}
\end{equation*}

We then apply the Born rule:

\begin{equation}
|p| = \langle s|P|s\rangle
\end{equation}

This geometrization process has the advantage of removing additional symbolic constructs such as \( |p|_q \) and of recasting probability algebra into a substantially Boolean form. Specifically, we have:

\begin{equation}\label{bool}
	\begin{cases}
		\neg P = I - P \\
		P \land Q = PQ = QP \\
		P \vee Q = P + Q - PQ
	\end{cases}
\end{equation}

Repeating this reasoning for \( N \) propositions \( p, q, r, \ldots \), we obtain that:

\begin{itemize}
	\item The semantic relationships among \( n \) propositions are captured by a state vector \( |s\rangle \), i.e., a direction, a one-dimensional subspace in \( \mathbb{R}^{2^n} \).\footnote{If the \( n \) propositions are not independent, the space may be of lower dimension, but we do not dwell on such details here.}
	\item A generic proposition \( p \), formed by combinations of the involved propositions, is associated with a suitable diagonal projector \( P \).
	\item The probability \( |p| \) is given by the Born rule \( \langle s|P|s\rangle \).
\end{itemize}

With this quick procedure, classical probability theory is given a pseudo-quantum geometric form in which Hilbert spaces are replaced by real vector spaces, and general Hermitian projectors are replaced by real diagonal projectors, or at least projectors that are all simultaneously diagonalizable via an appropriate change of basis.

\subsection{Geometric Representation of Relative Probability}

The geometrisation of CTP is founded on the probability values of the conjunctions $|p \land q|$, $|p \land \lnot q|$, $|\lnot p \land q|$, $|\lnot p \land \lnot q|$.  
We rewrite these expressions using symbols with serifs, $\mathcal{T}$ and $\mathcal{F}$, to denote absolute truth and falsity:
\begin{equation}
\{ |\mathcal{TT}|, |\mathcal{TF}|, |\mathcal{FT}|, |\mathcal{FF}| \}
\end{equation}

The system can then be extended by introducing the symbols $T$, $F$, $U$: "true and observable", "false and observable", "unobservable".  
One thus constructs the algebra of joint states:
\begin{equation*}
\{ |TT|, |TF|, |TU|, |FT|, |FF|, |FU|, |UT|, |UF|, |UU| \}
\end{equation*}

In this space, each proposition $p$ now contributes an $\mathbb{R}^3$ space ($T$, $F$, $U$), rather than an $\mathbb{R}^2$ space ($\mathcal{T}$, $\mathcal{F}$).  
Within this space, one can define the projector associated with the probability that $p$ is observable:
\begin{equation}
\bar{P} = \operatorname{diag}(1, 1, 1, 1, 1, 1, 0, 0, 0)
\end{equation}

And the projector associated with the probability that $p$ is true and observable:
\begin{equation}
\hat{P} = \operatorname{diag}(1, 1, 1, 0, 0, 0, 0, 0, 0)
\end{equation}

Thus, from \ref{relativeprobabilitypre}:
\begin{equation}\label{r3p}
[p] = \frac{\langle s | \hat{P} | s \rangle}{\langle s | \bar{P} | s \rangle}
\end{equation}

Equation \ref{r3p} is problematic in that it is non-linear; that is, there exists no operator $P$ such that:
\begin{equation}
[p] = \langle s | P | s \rangle
\end{equation}

This extension of the geometry therefore allows one to compute values such as $[p]$, but it loses the distinctive expressive properties—namely, the ability to reduce probabilistic calculation to a Boolean-like form as in \ref{bool}, in which operators can be directly multiplied and summed.  
In particular, the law of the excluded middle does not hold here in its most basic form:
\begin{equation}
\lnot P \neq I - P
\end{equation}

To resolve this algebraically in a satisfactory manner, it is necessary to \textit{linearise} the conditional probabilities; that is, to employ a space with the expressive capacity of $\mathbb{R}^3$, but with only two orthogonal directions—ensuring the law of the excluded middle—to which one can directly associate $[p]$ and $[\lnot p]$, along with exactly one additional degree of freedom.  
A space of this type exists: it is $\mathbb{C}^2$.

\subsection{Relative Probability and Quantum Mechanics}

\begin{figure}[H]
	\centering
	\includegraphics[width=0.7\linewidth]{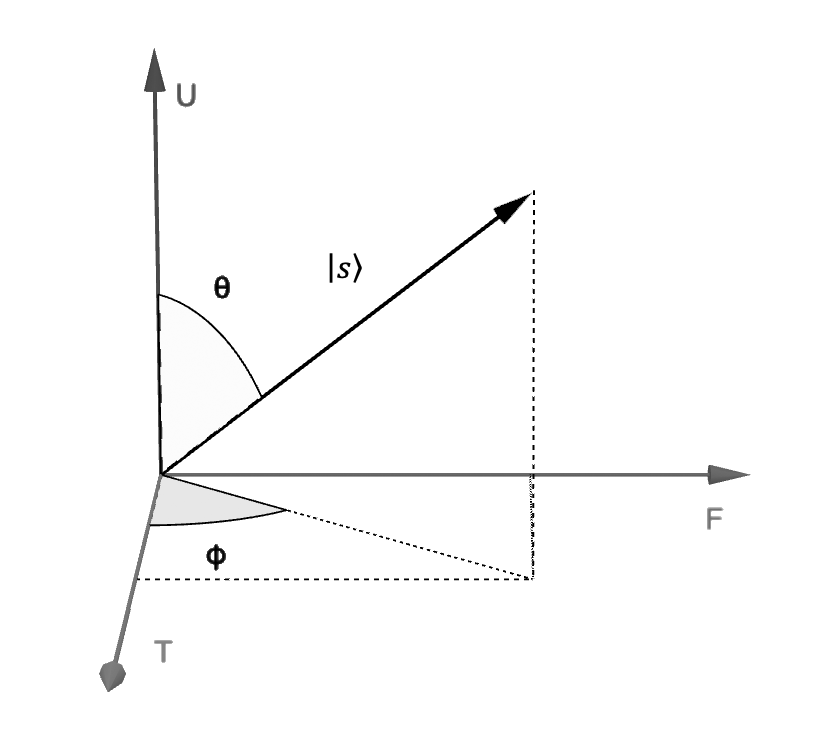}
	\caption[Didascalia]{}
	\label{fig:img1}
\end{figure}

The state of a single proposition $p$ can be represented as a direction in a space $\mathbb{R}^3$, whose three axes correspond to the conditions ``$p$ is true and observable,'' ``$p$ is false and observable,'' and ``$p$ is unobservable.'' Intermediate directions represent possible combinations of probabilities.  
Figure \ref{fig:img1} shows its representation in polar coordinates.

\begin{equation}
|s\rangle = 
\begin{pmatrix}
	\sin(\theta) \cos(\varphi) \\
	\sin(\theta) \sin(\varphi) \\
	\cos(\theta)
\end{pmatrix}
\end{equation}

The rotation that brings $|s\rangle$ onto the $TF$ plane is:

\begin{equation}
R = 
R_\varphi^\dagger \, R_\theta \, R_\varphi
\end{equation}

where:

\begin{equation}
\begin{cases}
R_\varphi = 
\begin{pmatrix}
	\sin(\varphi) & -\cos(\varphi) & 0 \\
	\cos(\varphi) & \sin(\varphi)  & 0 \\
	0            & 0              & 1
\end{pmatrix}
\\
R_\theta = 
\begin{pmatrix}
	1 & 0                & 0 \\
	0 & \sin(\theta)     & -\cos(\theta) \\
	0 & \cos(\theta)     & \sin(\theta)
\end{pmatrix}
\end{cases}
\end{equation}

And, by construction:

\begin{equation}
R |s\rangle = 
\begin{pmatrix}
	\cos(\varphi) \\
	\sin(\varphi) \\
	0
\end{pmatrix}
\end{equation}

It is therefore possible to directly extract the relative probability associated with $p$ using the projector:

\begin{equation}
P^* = 
\begin{pmatrix}
	1 & 0 & 0 \\
	0 & 0 & 0 \\
	0 & 0 & 0
\end{pmatrix}
\end{equation}

In the original basis, this operator takes the form:

\begin{equation}
P = R^\dagger P^* R = R_\varphi^\dagger R_\theta^\dagger R_\varphi P^* R_\varphi^\dagger R_\theta R_\phi
\end{equation}

The problem with this operator is that the $R_\theta$ components associated with unobservability are enclosed within a sandwich of components associated with observable probability $R_\varphi$; in other words, it is not possible to make $P$ independent of $\varphi$.

However, a radical simplification can be obtained through a process of \textit{complexification}, which maps the directions of $\mathbb{R}^3$ onto directions (up to a global phase) in $\mathbb{C}^2$.  
Consider the following map, similar to (but not identical to) the Bloch sphere:

\begin{equation*}
|s\rangle = 
\begin{pmatrix}
	\sin(\theta) \cos(\varphi) \\
	\sin(\theta) \sin(\varphi) \\
	\cos(\theta)
\end{pmatrix}
\quad \longrightarrow \quad
\begin{pmatrix}
	\cos(\varphi) \\
	e^{i \theta} \sin(\varphi)
\end{pmatrix}
= |s'\rangle
\end{equation*}

$|s'\rangle$ preserves the entire information content of $|s\rangle$ in a form that allows direct extraction of $[p]$ through a trivial projector:

\begin{equation}
P = 
\begin{pmatrix}
	1 & 0 \\
	0 & 0
\end{pmatrix}
\end{equation}

It is therefore possible to apply the Born rule and Boolean negation:

\begin{equation}
\begin{cases}
	[p] = \langle s | P | s \rangle \\
	\lnot P = I - P
\end{cases}
\end{equation}

A single proposition $p$ is thus represented by a quantum bit.  
Naturally, each possible point on the sphere in $R^3$ corresponds to a different epistemic condition—that is, to different $T$, $F$, $U$ probabilities—and therefore each observable defined on the qubit is associated with the preparation of the truth/falsehood measurement in an appropriate epistemic context.  

In particular, an arbitrary projector in $\mathbb{C}^2$ corresponds to testing a derived proposition $p'$, whose semantic content is defined by the chosen epistemic context—that is, by selecting a specific chain of phenomena through which the observer interrogates the system.  
From a logical perspective, $p'$ is a contextual proposition expressing \textit{``truth of $p$ under measurement context $U$,''} where $U$ is a unitary transformation of the canonical $TFU$ frame.  

The extension to $n$ propositions in $\mathbb{C}^{2^n}$ is entirely natural, observing that classical probability is subject to local tomography; this allows a reconstruction of the global state purely from the statistics of combined propositions—a direct analogue of tomographic locality in quantum mechanics~\cite{Hardy, Chiribella}.  

In conclusion, it is possible to assign a geometric structure to relative probability such that:

\begin{itemize}
	\item The semantic relations among $n$ propositions, including unobservable ones, are captured by a state vector $|S\rangle$—that is, a direction, a one-dimensional subspace—in $\mathbb{C}^{2^n}$.\footnote{In the case where the $n$ propositions are not independent, the space may have lower dimension, but such details are not pursued here.}
	\item A generic proposition $p$, formed by combinations of the involved propositions, is associated with an appropriate projector $P$.
	\item The probability $[p]$ is given by the Born rule: $\langle s | P | s \rangle$.
\end{itemize}

Or, more directly stated:

\begin{quote}
	The foundational principles of quantum mechanics define the algebra of semantic relations among propositions—even unobservable ones.
\end{quote}

\section{Conclusions}

The path followed in this work has shown how the presence of unobservability—in a physically grounded form—necessitates a non-classical probabilistic structure formally equivalent to that of quantum mechanics.

Brukner’s work demonstrates that the need for a scientific theory capable of handling random outcomes may stem from a strictly logical issue—the independence of propositions—rather than from any unusual ontological assumption. 

Moreover, these works suggest that QM offers an optimal framework for the treatment of unprovable propositions. This paper attempts to take a further step, showing that onto-epistemic ignorance necessarily entails a mathematical structure akin to that of QM—that is, not only does the independence of propositions lead to a probabilistic framework, but the specific form of onto-epistemic ignorance requires a formalism that, like QM, violates certain theorems of classical probability theory.

Within this interpretative scenario, Bell’s inequalities and the Wigner-d’Espagnat inequality are not, strictly speaking, violated—that is, QM does not violate logic. However, any empirical test of such inequalities inevitably tests a subtly stronger version of them:

\begin{equation}\label{WEs}
\{ \bar{A} \cap \bar{B} \} \cup \{ \overline{B^c} \cap \bar{C} \} \supseteq \{ \bar{A} \cap \bar{C} \}
\end{equation}

That is, the set of elements for which, relative to a concrete observer, properties \( A \) and \( B \) are both true and observable, together with the set for which \( \neg B \) and \( C \) are both true and observable, contains the set for which \( A \) and \( C \) are both true and observable.

And in this strengthened form, the proposition \ref{WEs} is false, since the set \( \{ \bar{A} \cap \bar{C} \} \) may contain an element whose membership in \( B \) is unobservable.

The conclusions reached here reinforce the suggestion—already present in Brukner—that quantum mechanics may be understood more as a framework dictated by logical necessity than by any particular ontology.  
In other words, that the epistemological problems associated with Schrödinger’s cat should not be addressed by focusing on the ontology of what lies inside the box, but rather on the box itself—as an idealised shield capable of disconnecting causal or phenomenal relations between its interior and exterior, thereby breaking the psycho-physical parallelism.  
This fact alone has significant consequences for the science conducted by the experimenter, in any situation in which they attempt to predict what will happen on the other side of the shield.

A thorough discussion of these deeper philosophical and interpretative aspects is inherently complex and divisive and is therefore left for future investigation.

\section*{Declarations}

This research received no specific grant from any funding agency in the public, commercial, or not-for-profit sectors.

\vspace{1em}
The author declares no conflict of interest.

\vspace{1em}
No experimental data were used or generated in the course of this study.

\vspace{1em}
The author is solely responsible for the content of this work.

\nocite{*}
\bibliography{Eng}

\begin{thebibliography}{24}%
\makeatletter
\providecommand \@ifxundefined [1]{%
 \@ifx{#1\undefined}
}%
\providecommand \@ifnum [1]{%
 \ifnum #1\expandafter \@firstoftwo
 \else \expandafter \@secondoftwo
 \fi
}%
\providecommand \@ifx [1]{%
 \ifx #1\expandafter \@firstoftwo
 \else \expandafter \@secondoftwo
 \fi
}%
\providecommand \natexlab [1]{#1}%
\providecommand \enquote  [1]{``#1''}%
\providecommand \bibnamefont  [1]{#1}%
\providecommand \bibfnamefont [1]{#1}%
\providecommand \citenamefont [1]{#1}%
\providecommand \href@noop [0]{\@secondoftwo}%
\providecommand \href [0]{\begingroup \@sanitize@url \@href}%
\providecommand \@href[1]{\@@startlink{#1}\@@href}%
\providecommand \@@href[1]{\endgroup#1\@@endlink}%
\providecommand \@sanitize@url [0]{\catcode `\\12\catcode `\$12\catcode
  `\&12\catcode `\#12\catcode `\^12\catcode `\_12\catcode `\%12\relax}%
\providecommand \@@startlink[1]{}%
\providecommand \@@endlink[0]{}%
\providecommand \url  [0]{\begingroup\@sanitize@url \@url }%
\providecommand \@url [1]{\endgroup\@href {#1}{\urlprefix }}%
\providecommand \urlprefix  [0]{URL }%
\providecommand \Eprint [0]{\href }%
\providecommand \doibase [0]{https://doi.org/}%
\providecommand \selectlanguage [0]{\@gobble}%
\providecommand \bibinfo  [0]{\@secondoftwo}%
\providecommand \bibfield  [0]{\@secondoftwo}%
\providecommand \translation [1]{[#1]}%
\providecommand \BibitemOpen [0]{}%
\providecommand \bibitemStop [0]{}%
\providecommand \bibitemNoStop [0]{.\EOS\space}%
\providecommand \EOS [0]{\spacefactor3000\relax}%
\providecommand \BibitemShut  [1]{\csname bibitem#1\endcsname}%
\let\auto@bib@innerbib\@empty
\bibitem [{\citenamefont {Bell}(1964)}]{Bell}%
  \BibitemOpen
  \bibfield  {author} {\bibinfo {author} {\bibfnamefont {J.~S.}\ \bibnamefont
  {Bell}},\ }\bibfield  {title} {\bibinfo {title} {On the einstein podolsky
  rosen paradox},\ }\href {https://doi.org/10.1103/PhysicsPhysiqueFizika.1.195}
  {\bibfield  {journal} {\bibinfo  {journal} {Physics Physique Fizika}\
  }\textbf {\bibinfo {volume} {1}},\ \bibinfo {pages} {195} (\bibinfo {year}
  {1964})}\BibitemShut {NoStop}%
\bibitem [{\citenamefont {Bell}(1981)}]{Bell2}%
  \BibitemOpen
  \bibfield  {author} {\bibinfo {author} {\bibfnamefont {J.~S.}\ \bibnamefont
  {Bell}},\ }\bibfield  {title} {\bibinfo {title} {Bertlmann's socks and the
  nature of reality},\ }\href@noop {} {\bibfield  {journal} {\bibinfo
  {journal} {Le Journal De Physique Colloques}\ }\textbf {\bibinfo {volume}
  {42}},\ \bibinfo {pages} {41} (\bibinfo {year} {1981})}\BibitemShut {NoStop}%
\bibitem [{\citenamefont {Bell}(2004)}]{Bell3}%
  \BibitemOpen
  \bibfield  {author} {\bibinfo {author} {\bibfnamefont {J.~S.}\ \bibnamefont
  {Bell}},\ }\href@noop {} {\emph {\bibinfo {title} {Speakable and unspeakable
  in quantum mechanics: Collected papers on quantum philosophy}}}\ (\bibinfo
  {publisher} {Cambridge university press},\ \bibinfo {year}
  {2004})\BibitemShut {NoStop}%
\bibitem [{\citenamefont {Aspect}\ \emph {et~al.}(1981)\citenamefont {Aspect},
  \citenamefont {Grangier},\ and\ \citenamefont {Roger}}]{Aspect}%
  \BibitemOpen
  \bibfield  {author} {\bibinfo {author} {\bibfnamefont {A.}~\bibnamefont
  {Aspect}}, \bibinfo {author} {\bibfnamefont {P.}~\bibnamefont {Grangier}},\
  and\ \bibinfo {author} {\bibfnamefont {G.}~\bibnamefont {Roger}},\ }\bibfield
   {title} {\bibinfo {title} {Experimental tests of realistic local theories
  via bell's theorem},\ }\href {https://doi.org/10.1103/PhysRevLett.47.460}
  {\bibfield  {journal} {\bibinfo  {journal} {Phys. Rev. Lett.}\ }\textbf
  {\bibinfo {volume} {47}},\ \bibinfo {pages} {460} (\bibinfo {year}
  {1981})}\BibitemShut {NoStop}%
\bibitem [{\citenamefont {Cirillo}(2015)}]{Cirillo}%
  \BibitemOpen
  \bibfield  {author} {\bibinfo {author} {\bibfnamefont {M.}~\bibnamefont
  {Cirillo}},\ }\href {https://doi.org/10.13140/RG.2.1.2198.9206} {\bibinfo
  {title} {Il teorema di bell raccontato a modo mio (... più o meno ...)}}
  (\bibinfo {year} {2015})\BibitemShut {NoStop}%
\bibitem [{\citenamefont {Bertlmann}(2020)}]{Bertlmann}%
  \BibitemOpen
  \bibfield  {author} {\bibinfo {author} {\bibfnamefont {R.~A.}\ \bibnamefont
  {Bertlmann}},\ }\bibfield  {title} {\bibinfo {title} {Real or not real that
  is the question...},\ }\href@noop {} {\bibfield  {journal} {\bibinfo
  {journal} {The European Physical Journal H}\ }\textbf {\bibinfo {volume}
  {45}},\ \bibinfo {pages} {205} (\bibinfo {year} {2020})}\BibitemShut
  {NoStop}%
\bibitem [{\citenamefont {Kolmogorov}(1950)}]{Kolmogorov}%
  \BibitemOpen
  \bibfield  {author} {\bibinfo {author} {\bibfnamefont {A.~N.}\ \bibnamefont
  {Kolmogorov}},\ }\href@noop {} {\emph {\bibinfo {title} {Foundations of the
  Theory of Probability}}}\ (\bibinfo  {publisher} {Chelsea Publishing
  Company},\ \bibinfo {address} {New York},\ \bibinfo {year} {1950})\ \bibinfo
  {note} {originally published in German as \textit{Grundbegriffe der
  Wahrscheinlichkeitsrechnung} (1933)}\BibitemShut {NoStop}%
\bibitem [{\citenamefont {Birkhoff}\ and\ \citenamefont
  {Neumann}(1936)}]{Birkhoff}%
  \BibitemOpen
  \bibfield  {author} {\bibinfo {author} {\bibfnamefont {G.}~\bibnamefont
  {Birkhoff}}\ and\ \bibinfo {author} {\bibfnamefont {J.~V.}\ \bibnamefont
  {Neumann}},\ }\bibfield  {title} {\bibinfo {title} {The logic of quantum
  mechanics},\ }\href {http://www.jstor.org/stable/1968621} {\bibfield
  {journal} {\bibinfo  {journal} {Annals of Mathematics}\ }\textbf {\bibinfo
  {volume} {37}},\ \bibinfo {pages} {823} (\bibinfo {year} {1936})}\BibitemShut
  {NoStop}%
\bibitem [{\citenamefont {VonNeumann}(1932)}]{VonNeumann}%
  \BibitemOpen
  \bibfield  {author} {\bibinfo {author} {\bibfnamefont {J.}~\bibnamefont
  {VonNeumann}},\ }\href {http://eudml.org/doc/203794} {\emph {\bibinfo {title}
  {Mathematische Grundlagen der Quantenmechanik}}}\ (\bibinfo  {publisher}
  {Springer},\ \bibinfo {year} {1932})\BibitemShut {NoStop}%
\bibitem [{\citenamefont {Benioff}(1976)}]{Benioff}%
  \BibitemOpen
  \bibfield  {author} {\bibinfo {author} {\bibfnamefont {P.~A.}\ \bibnamefont
  {Benioff}},\ }\bibfield  {title} {\bibinfo {title} {Models of
  zermelo-fraenkel set theory as carriers for the mathematics of physics. i},\
  }\href {https://doi.org/10.1063/1.522984} {\bibfield  {journal} {\bibinfo
  {journal} {Journal of Mathematical Physics}\ }\textbf {\bibinfo {volume}
  {17}},\ \bibinfo {pages} {618} (\bibinfo {year} {1976})}\BibitemShut
  {NoStop}%
\bibitem [{\citenamefont {Wheeler}(1974)}]{Wheeler}%
  \BibitemOpen
  \bibfield  {author} {\bibinfo {author} {\bibfnamefont {J.}~\bibnamefont
  {Wheeler}},\ }\href@noop {} {\bibinfo {title} {Add “participant” to
  “undecidable propositions” to arrive at physics}},\ \bibinfo
  {howpublished} {https://jawarchive.files.wordpress.com/2012/03/twa-1974.pdf}
  (\bibinfo {year} {1974})\BibitemShut {NoStop}%
\bibitem [{\citenamefont {Poletti}(2022)}]{Poletti}%
  \BibitemOpen
  \bibfield  {author} {\bibinfo {author} {\bibfnamefont {M.}~\bibnamefont
  {Poletti}},\ }\bibfield  {title} {\bibinfo {title} {On the strangeness of
  quantum mechanics},\ }\href {https://doi.org/10.1007/s10701-022-00582-w}
  {\bibfield  {journal} {\bibinfo  {journal} {Foundations of Physics}\ }\textbf
  {\bibinfo {volume} {52}},\ \bibinfo {pages} {1} (\bibinfo {year}
  {2022})}\BibitemShut {NoStop}%
\bibitem [{\citenamefont {Poletti}(2023{\natexlab{a}})}]{Poletti2}%
  \BibitemOpen
  \bibfield  {author} {\bibinfo {author} {\bibfnamefont {M.}~\bibnamefont
  {Poletti}},\ }\bibfield  {title} {\bibinfo {title} {On the strangeness of
  quantum probabilities},\ }\href {https://doi.org/10.1007/s40509-023-00299-z}
  {\bibfield  {journal} {\bibinfo  {journal} {Quantum Studies: Mathematics and
  Foundations}\ }\textbf {\bibinfo {volume} {10}},\ \bibinfo {pages} {343}
  (\bibinfo {year} {2023}{\natexlab{a}})}\BibitemShut {NoStop}%
\bibitem [{\citenamefont {Poletti}(2023{\natexlab{b}})}]{Poletti3}%
  \BibitemOpen
  \bibfield  {author} {\bibinfo {author} {\bibfnamefont {M.}~\bibnamefont
  {Poletti}},\ }\bibfield  {title} {\bibinfo {title} {Quanta iff
  discreteness},\ }\href@noop {} {\bibfield  {journal} {\bibinfo  {journal}
  {International Journal of Quantum Foundations}\ }\textbf {\bibinfo {volume}
  {9}},\ \bibinfo {pages} {174} (\bibinfo {year}
  {2023}{\natexlab{b}})}\BibitemShut {NoStop}%
\bibitem [{\citenamefont {Poletti}(2023{\natexlab{c}})}]{Poletti4}%
  \BibitemOpen
  \bibfield  {author} {\bibinfo {author} {\bibfnamefont {M.}~\bibnamefont
  {Poletti}},\ }\href {https://arxiv.org/abs/2308.03341} {\bibinfo {title}
  {Bertlmann's socks from a viennese perspective}} (\bibinfo {year}
  {2023}{\natexlab{c}}),\ \Eprint {https://arxiv.org/abs/2308.03341}
  {arXiv:2308.03341 [physics.hist-ph]} \BibitemShut {NoStop}%
\bibitem [{\citenamefont {Szangolies}(2018)}]{Szangolies}%
  \BibitemOpen
  \bibfield  {author} {\bibinfo {author} {\bibfnamefont {J.}~\bibnamefont
  {Szangolies}},\ }\bibfield  {title} {\bibinfo {title} {Epistemic horizons and
  the foundations of quantum mechanics},\ }\href
  {https://doi.org/10.1007/s10701-018-0221-9} {\bibfield  {journal} {\bibinfo
  {journal} {Foundations of Physics}\ }\textbf {\bibinfo {volume} {48}},\
  \bibinfo {pages} {1669–} (\bibinfo {year} {2018})}\BibitemShut {NoStop}%
\bibitem [{\citenamefont {Landsman}(2021)}]{Landsman}%
  \BibitemOpen
  \bibfield  {author} {\bibinfo {author} {\bibfnamefont {K.}~\bibnamefont
  {Landsman}},\ }\href {https://doi.org/10.1007/978-3-030-70354-7_3} {\bibinfo
  {title} {Indeterminism and undecidability}} (\bibinfo {year}
  {2021})\BibitemShut {NoStop}%
\bibitem [{\citenamefont {Paterek}\ \emph {et~al.}(2010)\citenamefont
  {Paterek}, \citenamefont {Kofler}, \citenamefont {Prevedel}, \citenamefont
  {Klimek}, \citenamefont {Aspelmeyer}, \citenamefont {Zeilinger},\ and\
  \citenamefont {{Brukner, \v{C}}}}]{Brukner}%
  \BibitemOpen
  \bibfield  {author} {\bibinfo {author} {\bibfnamefont {T.}~\bibnamefont
  {Paterek}}, \bibinfo {author} {\bibfnamefont {J.}~\bibnamefont {Kofler}},
  \bibinfo {author} {\bibfnamefont {R.}~\bibnamefont {Prevedel}}, \bibinfo
  {author} {\bibfnamefont {P.}~\bibnamefont {Klimek}}, \bibinfo {author}
  {\bibfnamefont {M.}~\bibnamefont {Aspelmeyer}}, \bibinfo {author}
  {\bibfnamefont {A.}~\bibnamefont {Zeilinger}},\ and\ \bibinfo {author}
  {\bibnamefont {{Brukner, \v{C}}}},\ }\bibfield  {title} {\bibinfo {title}
  {Logical independence and quantum randomness},\ }\href
  {https://doi.org/10.1088/1367-2630/12/1/013019} {\bibfield  {journal}
  {\bibinfo  {journal} {New Journal of Physics}\ }\textbf {\bibinfo {volume}
  {12}},\ \bibinfo {pages} {013019} (\bibinfo {year} {2010})}\BibitemShut
  {NoStop}%
\bibitem [{\citenamefont {Brukner}(2008)}]{Brukner2}%
  \BibitemOpen
  \bibfield  {author} {\bibinfo {author} {\bibfnamefont {{\v{C}}.}~\bibnamefont
  {Brukner}},\ }\bibfield  {title} {\bibinfo {title} {Quantum experiments can
  test mathematical undecidability},\ }in\ \href@noop {} {\emph {\bibinfo
  {booktitle} {Unconventional Computing}}},\ \bibinfo {editor} {edited by\
  \bibinfo {editor} {\bibfnamefont {C.~S.}\ \bibnamefont {Calude}}, \bibinfo
  {editor} {\bibfnamefont {J.~F.}\ \bibnamefont {Costa}}, \bibinfo {editor}
  {\bibfnamefont {R.}~\bibnamefont {Freund}}, \bibinfo {editor} {\bibfnamefont
  {M.}~\bibnamefont {Oswald}},\ and\ \bibinfo {editor} {\bibfnamefont
  {G.}~\bibnamefont {Rozenberg}}}\ (\bibinfo  {publisher} {Springer Berlin
  Heidelberg},\ \bibinfo {address} {Berlin, Heidelberg},\ \bibinfo {year}
  {2008})\ pp.\ \bibinfo {pages} {1--5}\BibitemShut {NoStop}%
\bibitem [{\citenamefont {Časlav Brukner}(2009)}]{Brukner3}%
  \BibitemOpen
  \bibfield  {author} {\bibinfo {author} {\bibnamefont {Časlav Brukner}},\
  }\bibfield  {title} {\bibinfo {title} {Quantum complementarity and logical
  indeterminacy},\ }\href {https://doi.org/10.1007/s11047-009-9118-z}
  {\bibfield  {journal} {\bibinfo  {journal} {Natural Computing}\ }\textbf
  {\bibinfo {volume} {8}},\ \bibinfo {pages} {449} (\bibinfo {year}
  {2009})}\BibitemShut {NoStop}%
\bibitem [{\citenamefont {VonNeumann}(2008)}]{VonNeumann2008}%
  \BibitemOpen
  \bibfield  {author} {\bibinfo {author} {\bibfnamefont {J.}~\bibnamefont
  {VonNeumann}},\ }\href
  {https://archive.org/details/mathematical-foundations-of-quantum-mechanics-2018}
  {\emph {\bibinfo {title} {Mathematical Foundations of Quantum Mechanics, New
  Edition}}}\ (\bibinfo  {publisher} {Princeton University Press},\ \bibinfo
  {year} {2008})\BibitemShut {NoStop}%
\bibitem [{\citenamefont {Kant}(1781)}]{Kant}%
  \BibitemOpen
  \bibfield  {author} {\bibinfo {author} {\bibfnamefont {I.}~\bibnamefont
  {Kant}},\ }\href@noop {} {\emph {\bibinfo {title} {Critique of Pure
  Reason}}}\ (\bibinfo  {publisher} {Macmillan and Co.},\ \bibinfo {address}
  {London},\ \bibinfo {year} {1781})\ \bibinfo {note} {first edition (A
  edition); translated by Norman Kemp Smith, 1929}\BibitemShut {NoStop}%
\bibitem [{\citenamefont {Hardy}(2001)}]{Hardy}%
  \BibitemOpen
  \bibfield  {author} {\bibinfo {author} {\bibfnamefont {L.}~\bibnamefont
  {Hardy}},\ }\href {https://arxiv.org/abs/quant-ph/0101012} {\bibinfo {title}
  {Quantum theory from five reasonable axioms}} (\bibinfo {year} {2001}),\
  \Eprint {https://arxiv.org/abs/quant-ph/0101012} {arXiv:quant-ph/0101012
  [quant-ph]} \BibitemShut {NoStop}%
\bibitem [{\citenamefont {Chiribella}\ \emph {et~al.}(2011)\citenamefont
  {Chiribella}, \citenamefont {D'Ariano},\ and\ \citenamefont
  {Perinotti}}]{Chiribella}%
  \BibitemOpen
  \bibfield  {author} {\bibinfo {author} {\bibfnamefont {G.}~\bibnamefont
  {Chiribella}}, \bibinfo {author} {\bibfnamefont {G.~M.}\ \bibnamefont
  {D'Ariano}},\ and\ \bibinfo {author} {\bibfnamefont {P.}~\bibnamefont
  {Perinotti}},\ }\bibfield  {title} {\bibinfo {title} {Informational
  derivation of quantum theory},\ }\href
  {https://doi.org/10.1103/PhysRevA.84.012311} {\bibfield  {journal} {\bibinfo
  {journal} {Phys. Rev. A}\ }\textbf {\bibinfo {volume} {84}},\ \bibinfo
  {pages} {012311} (\bibinfo {year} {2011})}\BibitemShut {NoStop}%
\end{thebibliography}%

\end{document}